\documentclass[showpacs,showkeys,preprintnumbers,amsmath,amssymb,latexsym,onecolumn]{revtex4}

\usepackage{graphicx}
\usepackage{dcolumn}
\usepackage{bm}
\usepackage{epsfig}
\usepackage{natbib}

\def\ut#1{\mathop{\vtop{\ialign{##\crcr
     $\hfil\displaystyle{#1}\hfil$\crcr\noalign
     {\kern1pt\nointerlineskip}\hbox{$\hfil\sim\hfil$}\crcr
     \noalign{\kern1pt}}}}}

\def\undersymbol#1#2{\mathop{\vtop{\ialign{##\crcr
     $\hfil\displaystyle{#2}\hfil$\crcr\noalign
     {\kern1pt\nointerlineskip}\hbox{$\hfil#1\hfil$}\crcr
     \noalign{\kern1pt}}}}}
\def\arcsec{^{\prime\prime}}

\begin{document}
\preprint{}
\title{Emission of gravitational waves from binary systems in the galactic center and diffraction by star clusters}

 \author{P. Longo}
 \email{longo@le.infn.it}
 \affiliation{Dipartimento di Fisica,
 Universit\`a di Lecce, and INFN, Sezione di Lecce, Via Arnesano,
 CP 193, I-73100 Lecce, Italy}

\author{G. Congedo}
 \email{congedo@le.infn.it}
 \affiliation{Dipartimento di Fisica,
 Universit\`a di Lecce, and INFN, Sezione di Lecce, Via Arnesano,
 CP 193, I-73100 Lecce, Italy}

\author{A. A. Nucita}
 \email{nucita@le.infn.it}
 \affiliation{Dipartimento di Fisica,
 Universit\`a di Lecce, and INFN, Sezione di Lecce, Via Arnesano,
 CP 193, I-73100 Lecce, Italy}
\date{\today}

\author{F. De Paolis}
 \email{depaolis@le.infn.it}
 \affiliation{Dipartimento di Fisica,
 Universit\`a di Lecce, and INFN, Sezione di Lecce, Via Arnesano,
 CP 193, I-73100 Lecce, Italy}

\author{G. Ingrosso}
 \email{ingrosso@le.infn.it}
 \affiliation{Dipartimento di Fisica,
 Universit\`a di Lecce, and INFN, Sezione di Lecce, Via Arnesano,
 CP 193, I-73100 Lecce, Italy}


\begin{abstract}

Binary systems of compact objects are strong emitters of
gravitational waves whose amplitude depends on the binary orbital
parameters as the component mass, the orbital semi-major axis and
eccentricity. Here, in addition to the famous Hulse-Taylor binary
system, we have studied the possibility to detect the
gravitational wave signal emitted by binary systems at the center
of our galaxy. In particular, recent infrared observation of the
galactic center have revealed the existence of a cluster of stars
each of which appears to orbit the central black hole in $SgrA^*$.
For the stars labelled as S2 and S14, we have studied the emitted
spectrum of gravitational wave and compare it with the sensitivity
threshold of space-based interferometers like Lisa and Astrod.
Furthermore, following recent observations, we have considered the
possibility that $SgrA^*$ is actually a binary system of massive
black holes and calculated the emission spectrum as a function of
the system parameters. The diffraction pattern of gravitational
waves emitted by a binary system by a cluster of stars has been
also analyzed. We remark that this is only a
preliminary-theoretical work than can acquire more interest in
view of the next-coming gravitational wave astronomy era.
\end{abstract}

 \keywords{}
\maketitle


\section{introduction}
The nucleus of our Galaxy is only $8.5$ kpc far from Earth.
Consequently, it offers a unique possibility to study some
physical processes with a level of details that will never be
reached in external galaxies or active galactic nuclei. Thus, a
consistent theoretical picture of the observed physical phenomena
may allow improving not only our understanding of the galactic
structure but also our general view of other galactic nuclei.

The nature of the dark object at the  Galactic Center is still
unclear although a super massive black hole (SMBH) of about a few
million solar masses seems to be the most viable scenario. The
corresponding Schwarzschild radius of such a black hole is
$R_s=2GM/c^2\simeq 10^{11}$ cm which, at the distance of $\sim
8.5$ kpc, corresponds to an angular size of a few $\mu$as. Since
no present telescope has an angular resolution comparable with
that required, only indirect proofs of the existence of a SMBH at
the Galactic Center can be obtained.

A proof of the existence of a SMBH and its association with Sgr
A$^*$ lies in assessment of the mass distribution in the central
few parsec of the Galaxy \footnote{Other evidences of the presence
of a SMBH at the galactic center may be provided by future
retro-lensing observations \cite{depaoliss2,miraggi} }. If the
gravitational force is the dominant force acting in the vicinity
of the SMBH, the velocities and the orbits of nearby stars
strongly depend on the mass of central black hole. Hence, a number
of efforts have been spent in order to map the galactic center
region.

Recent ESO and Keck infrared observations have revealed the
existence of a cluster of stars (see Figure \ref{s2orbit}) in the
vicinity of the Galactic Center ($< 1\arcsec $)
\cite{Genzel03,Schoedel03,Ghez03,Ghez04,Ghez05}. In particular,
Ghez et al. (\citeyear{Ghez03}) have reported on observations of a
star orbiting close to the galactic center massive black hole. The
star, which has been labelled as S2, with mass $M_{\rm S2}\simeq
15$ M$_{\odot}$, appears to be a main sequence star, orbiting the
black hole in SgrA$^*$ with a Keplerian period of $\simeq 15$ yrs.
This has allowed Ghez et al. \citeyear{Ghez05} to estimate a mass
of $M_{\rm SgrA^*}\simeq 3.67\times 10^6 ~M_{\odot}$ within
$4.87\times 10^{-3}$ pc. The orbital parameters for the binary
system $S2-SgrA^{*}$  are given in Table \ref{S2tab}.

The relatively short periastron distance of both the S2 and S14
stars encourages the attempt of an observational campaign to look
for genuine relativistic effects like the emission of
gravitational waves (GWs). Starting from the Einstein equations
and quadrupole radiation formula \cite{pet}, it is possible to
evaluate, in the weak field regime,  the expected GW signal by the
binary system SgrA$^*$-S2 (and SgrA$^*$-S14) and compare the
emitted GW spectrum with the threshold of next space-based
interferometers like Lisa and Astrod. Moreover, according to
recent observations, we have considered the possibility that
$SgrA^*$ is actually a binary system of black holes orbiting the
common center of mass.

In addition to the emission of gravitational waves form binary
systems in the galactic center, we have also considered the effect
of diffraction on a cluster of stars surrounding the central
massive black hole. In fact, since in the linearized theory of
General Relativity a gravitational wave acts as an ordinary
electromagnetic wave, we expect that, under particular condition,
it suffers diffractive effects while interacting with stars.
Hence, by considering each star as a circular slit and applying
the well known theory of wave diffraction, we can evaluate the
expected diffraction patterns on the observer plane.

The paper is structured as follows: in Section II, we study the
emission of gravitational waves from binary systems in the
galactic center and compare the emitted spectrum with Lisa and
Astrod sensitivity curves. Section II A, is devoted to the
SgrA$^*$-S2 (and SgrA$^*$-S14) binary systems, while in Section II
B we focus on the possibility that the galactic center hosts a
massive black hole binary. In Section III A, we study the
diffraction of gravitational waves by a stellar cluster possibly
surrounding the SgrA$^*$ black hole. Here, by using Montecarlo
techniques, we simulate the distribution of stars (each of which
acts as a circular hole) at the galactic center and evaluate the
diffraction patterns as observed from Earth.

\section{Emission of GWs by binary systems}

The Theory of General Relativity predicts that a system of moving
bodies emits gravitational waves. In particular, two stars on
circular orbit release GWs characterized by a frequency $\omega_n$
which is twice the orbital one ($\omega _k$). Stars on an elliptic
orbit (with semi-major axis $a$ and eccentricity $e$) emit GWs at
frequencies $\omega_n = n \omega_k$ (with $n=1,2,3...$). In this
case, the Fourier analysis gives the following expression for the
released GW power in the n-th harmonic \citep{pet}

\begin{equation}\label{potenza}
\frac{dE}{dt}(n)=\frac{32}{5}\frac{G^4}{c^5}\frac{m_1^2m_2^2}{a^5}(m_1+m_2)g(n,e)
\end{equation}
where
\begin{equation}\label{bessel}
\begin{array}{cc}
g(n,e)=\displaystyle{\frac{n^4}{32}}\{[J_{n-2}(ne)-2eJ_{n-1}(ne)+\frac{2}{n}J_n(ne)+
2eJ_{n+1}(ne)-J_{n+2}(ne)]^2+\\
(1-e^2)\left[J_{n-2}(ne)-2J_n(ne)+J_{n+2}(ne)\right]^2
+\displaystyle{\frac{4}{3n^2}}\left[J_n(ne)\right]^2\},\\
\end{array}
\end{equation}
where $J_n$ (with $n$ integer) is the usual Bessel function of $n$
order.

Peters and Mathews (\cite{pet}), found that the orbit averaged
power emitted in GW by a binary system with eccentricity $e$ in
the {\it n}th harmonic of the orbital frequency is
\begin{equation}\label{ampiezza}
h(\omega_{n})\simeq\frac{2}{\pi\omega_{n}D}\sqrt{\frac{G}{c^{3}}\frac{dE(n,e)}{dt}},
\end{equation}
where $D$ is the distance of the gravitational wave source from
Earth.

\begin{figure}[h]
\begin{center}
\hbox{\hspace{2cm}
\psfig{figure=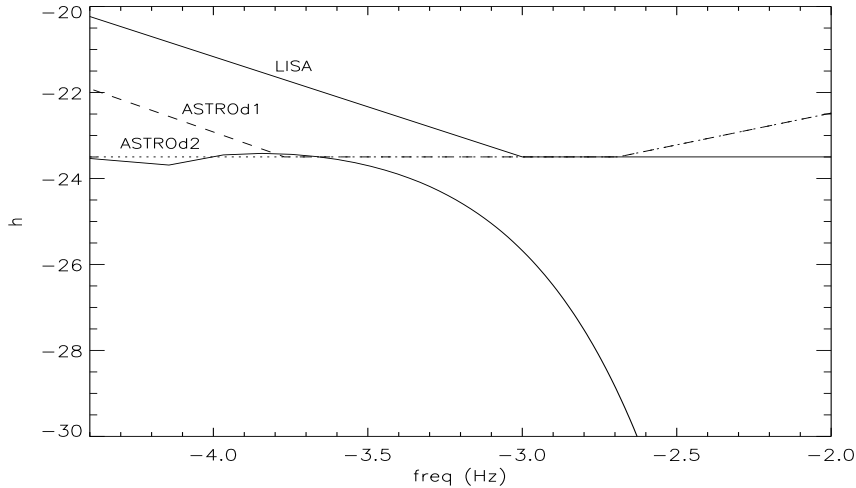,height=7.0cm,width=12.0cm}
} \vspace{-1.0cm}
\end{center}
\caption{\footnotesize {The gravitational wave spectrum expected
from the Hulse-Taylor binary pulsar is shown and compared with
Lisa and  Astrod thresholds. The sensitivity curves have been
obtaine for an integration time of 5 yrs.}} \label{speht}
\end{figure}

As an example, we may apply the previous relations in order to
determine the expected gravitational wave signal form the well
known Hulse-Taylor binary pulsar ($PSR 1913+16$) \cite{hulse}. In
Table \ref{HTtab}, we give the main orbital parameters of $PSR
1913+16$ while the expected GW spectrum is shown in Figure
\ref{speht} where it is compared with the Lisa and Astrod
sensitivity curves (obtained with an integration time of 5 yrs).
Inspection of that Figure shows that the gravitational wave signal
from the binary pulsar $PSR 1913+16$ is close to the threshold of
Astrod but is rather far from the possibility of being observed by
Lisa.
\begin{table}
\begin{center}
\begin{tabular}{|c|c|}
 \hline
Parameters & PSR 1913+16 \\
\hline
Distance (Kpc)  & 7.3 \\
Longest semiaxis $a$ (cm) & $1.9\times 10^{11}$ \\
Eccentricity $e$ & 0.617130 \\
Orbital period (hr)  & 7.7519 \\
$m_1+m_2(M_{\odot})$  & 2.82837 \\
\hline
\end{tabular}
\end{center}
\caption{\footnotesize{Main orbital parameters and distance form
Earth of the binary pulsar PSR 1913+16.}} \label{HTtab}
\end{table}

\subsection{Gravitational waves from SgrA$^*$-S2 (and SgrA$^*$-S14) binary system}

\begin{figure}[h]
\begin{center}
\hbox{\hspace{4.5cm}
\psfig{figure=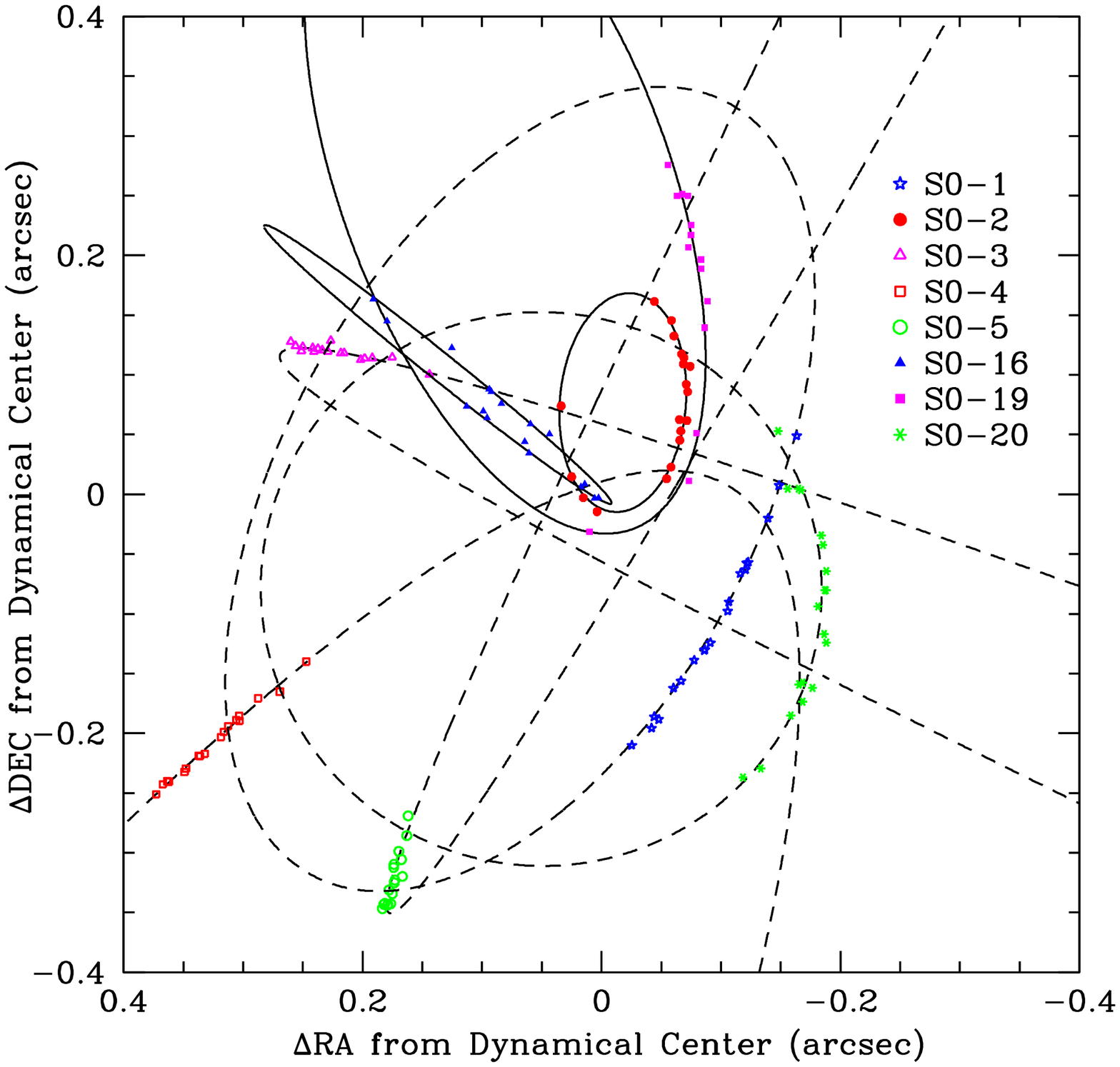,height=8.56cm,width=8.56cm}
} \vspace{-1.0cm}
\end{center}
\caption{\footnotesize {The orbits followed by few stars around
SgrA*  are shown. Data are taken from Ghez et al.
\citeyear{Ghez03}.)}} \label{s2orbit}
\end{figure}

By using eqs. (\ref{potenza})-(\ref{ampiezza}) it is
straightforward to calculate the GW spectrum for binary systems
hosted at the galactic center. In fact, as recent infrared
observations have shown, a massive black hole (with mass $M_{\rm
SgrA^*}\simeq 3.67\times 10^6 ~M_{\odot}$ ) is surrounded by a
cluster of at least 40 stars closer than $1.2\arcsec$ (see Figure
\ref{s2orbit}). In particular the star labelled as $S2$, whose
orbital parameters are given in Table \ref{S2tab}, is the most
interesting. Since its orbit will be entirely observed within few
years leading to test some relativistic effects, such as the star
periastron precession, it will likely allow the acquisition of
important information about the central black hole mass, its spin
and the mass distribution around it (see Bini et al.
\citeyear{bini2005}, De Paolis et al. \citeyear{depaolis2005} and
references therein).
\begin{table}
\begin{center}
\begin{tabular}{|c|c|}
\hline
Parameters & $S2-SgrA^*$ \\
\hline
Distance (Kpc) & 8.5 \\
$M_{BH}~(M_{\odot})$    & $3.67\times 10^6$ \\
$M_{S2}~(M_{\odot})$    & 15 \\
Longest semiaxis $a$ (pc)   & $4.87 \times 10^{-3}$\\
Eccentricity $e$        & 0.87 \\
Orbital period (year)  & 15.78 \\
\hline
\end{tabular}
\end{center}
\caption{\footnotesize{Parameters of the binary system
$S2-SgrA^{*}$.}} \label{S2tab}
\end{table}
\begin{figure}[h]
\begin{center}
\hbox{\hspace{2cm}
\psfig{figure=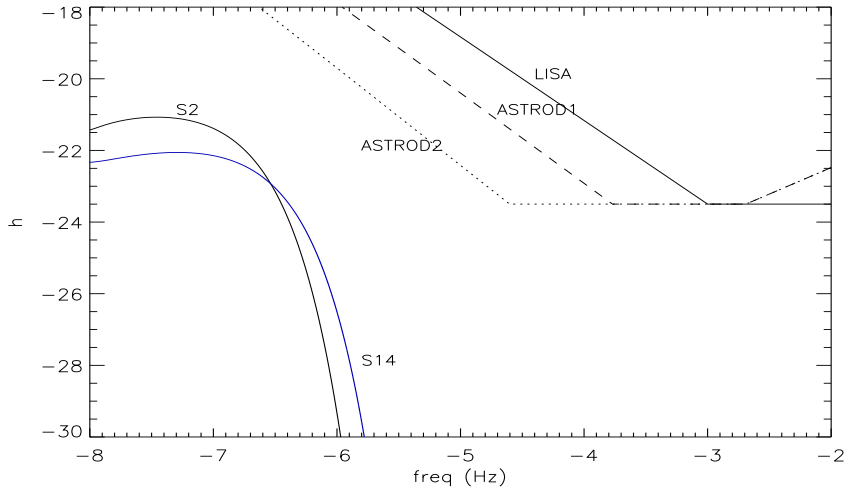,height=7.0cm,width=12.0cm}
} \vspace{-1.0cm}
\end{center}
\caption{\footnotesize {Emission spectrum of gravitational waves
from the binary systems $S2-SgrA^{*}$ and $S14-SgrA^{*}$ together
with the threshold curves of LISA, ASTROD1 and ASTROD2
interferometers.}} \label{spes2s14}
\end{figure}
In Figure \ref{spes2s14} the GW spectrum of the $S14$ star (with
$e=0.97$ and $a=15.1\cdot 10^{-3}~\rm{pc}$) is also given. As it
is clear from the figure, none of these binary systems is a GW
source detectable by the next generation of gravitational wave
space-based interferometers.

\subsection{Gravitational waves from a massive binary black hole in the galactic center}
\begin{figure}[h]
\begin{center}
\hbox{\hspace{2cm}
\psfig{figure=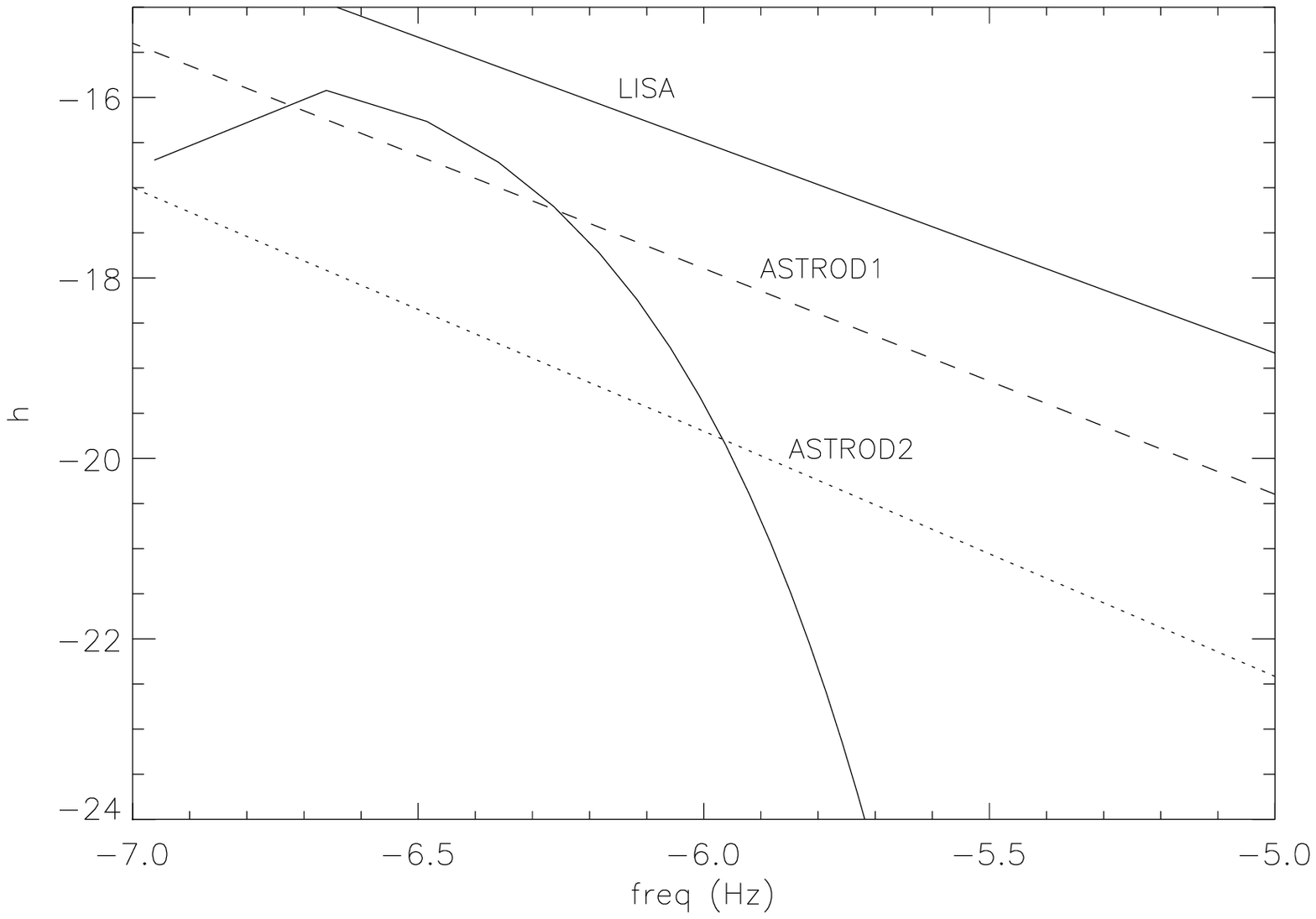,height=7.0cm,width=12.0cm}
} \vspace{-1.0cm}
\end{center}
\caption{\footnotesize {The gravitational wave spectrum from a
binary system of two black holes with mass ratio $q=0.01$ and
eccenticity $e=0.2$. The threshold curves of LISA, ASTROD1 and
ASTROD2 interferometers are also shown (with integration time of 5
yrs).}} \label{q001e02}
\end{figure}
\begin{figure}[h]
\begin{center}
\hbox{\hspace{2cm}
\psfig{figure=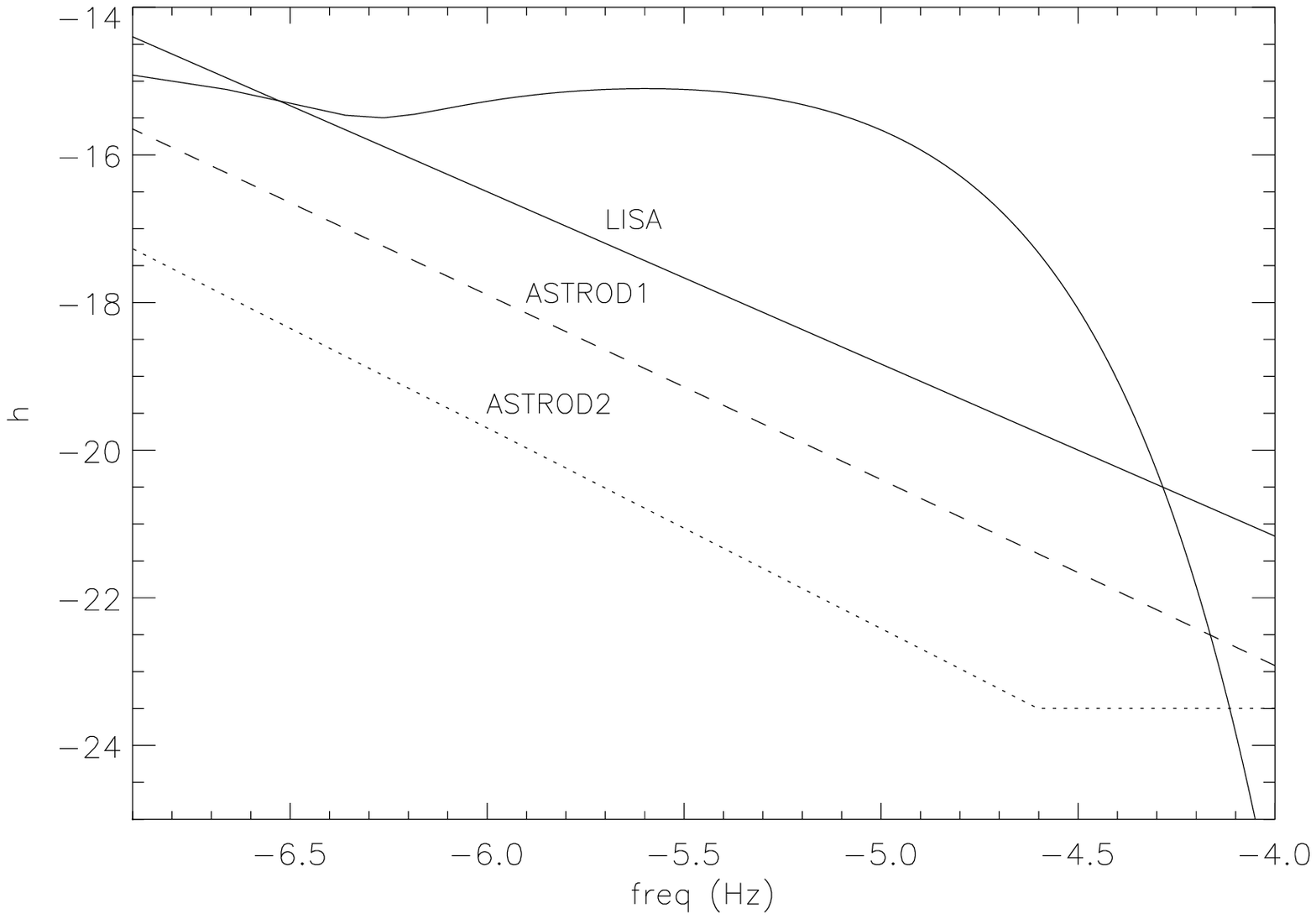,height=7.0cm,width=12.0cm}
} \vspace{-1.0cm}
\end{center}
\caption{\footnotesize {The gravitational wave spectrum from a
binary system of two black holes with mass ratio $q=0.9$ and
eccenticity $e=0.9$. The threshold curves of LISA, ASTROD1 and
ASTROD2 interferometers are also shown (with integration time of 5
yrs).}} \label{q09e09}
\end{figure}


The nucleus of our Galaxy hosts a dark object whose nature is
still unclear although a super massive black hole (SMBH) of
$M_{BH}=3.67\times 10^6~M_\odot$ seems to be the most viable
scenario. However, recent observations of the galactic center
region in the radio band  have pointed out a possible periodic
flux variation with a period of about $106$ days \citep{lom}. This
periodicity may be the consequence of a Doppler shift modulation
of the radio signal due to the orbital motion of a  massive black
hole binary system. For a system of this kind, eq. (\ref{potenza})
becomes
\begin{equation}\label{potenzabbh}
\frac{dE}{dt}(n)=\frac{32}{5}\frac{G^4}{c^5}\frac{M_{BH}^5}{a^5}\frac{q^2}{(1+q)^4}g(n,e)
\end{equation}
where $M_{BH}=3.67\times 10^6~M_\odot$ is the $SgrA^{*}$ total
mass. The obtained GW spectrum is shown in Figure \ref{q001e02}
and Figure \ref{q09e09}, for different values of eccentricity and
mass ratio $q=m1/m2$ (here $m_1$ and $m_2$ are the black hole
masses). As one can see, the next generation of space-based
interferometers should be able to detect GW from this kind of
object at the galactic center and confirm, or not, the existence
of a massive binary black hole.

\section{Diffraction of gravitational waves by a stellar cluster}

In 1999 Ruffa \citep{ruffa} has studied the diffraction of GWs by
a massive black hole at the galactic center. In that work the
black hole is described as a circular ring with thickness between
$R_E$ and $R_E+dR_E$ ($R_E$ is the black hole Einstein radius).
Ruffa showed that this phenomenon should amplify the GW amplitude
by a factor of about 130. A more realistic calculation
\citep{depaolis} has shown that the GW amplification factor is
actually smaller than Ruffa's one, due to the small chance of
having source, lens (black hole) and observer (Earth) aligned.

Here, we consider the GW diffraction by a star cluster adopting
the following simplifying assumptions:
\begin{enumerate}
    \item the stars are ``frozen'' in the cluster, that is, their movement is neglected;
    \item the cluster has a bi-dimensional distribution;
    \item each star is assumed to behave as a circular hole (slit) with respect to GW diffraction (due to the Babinet principle).
\end{enumerate}
The first hypothesis allows us to neglect the dynamical effects of
the system on the diffraction pattern. The second one is necessary
due to the long computational time needed for the simulations.

In the following two paragraph we introduce the main physical and
geometrical cluster parameters. Hence, we derive a relation which
makes some restrictions on these parameters in order to get GW
diffraction.

\subsection{Diffraction conditions on the cluster parameters}

Starting from the hypotheses discussed above, we have studied the
diffraction of GW emitted by a monochromatic GW source behind the
cluster (in the linear approximation of the Einstein equations).
First of all, we have assumed that the cluster mass density
distribution follows a Plummer model (see Figure \ref{Stellar
distribution} for more details).

\begin{figure}[h]
\begin{center}
\hbox{\hspace{1.5cm}
\psfig{figure=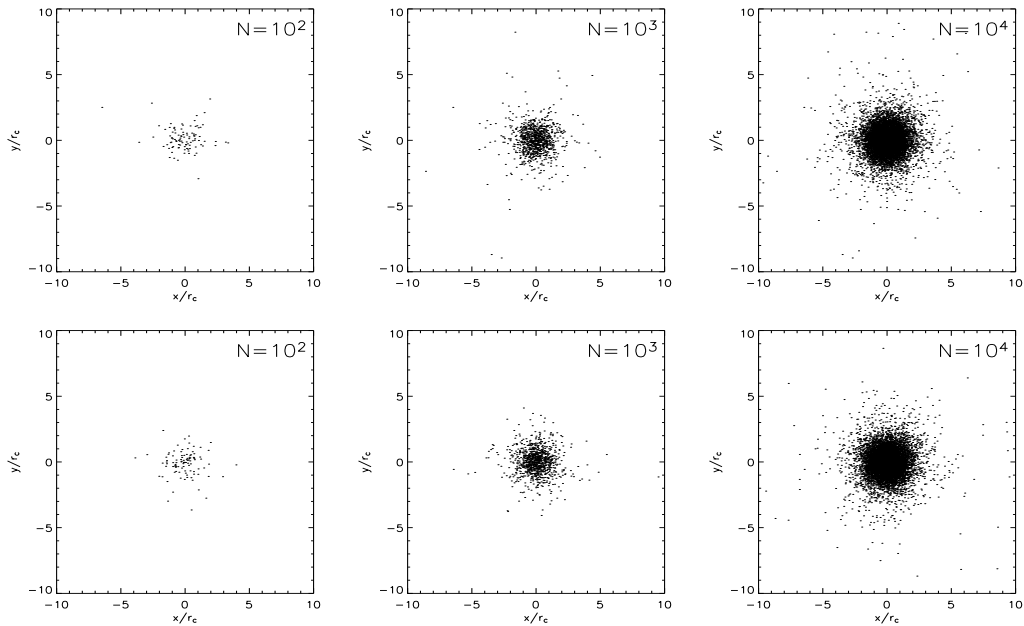,height=8.56cm,width=14.0cm}
} \vspace{-1.0cm}
\end{center}
\caption{\footnotesize {The stellar distributions obtained for a
number of stars of $N=10^2,~10^3$ and $10^4$, respectively. The
$x$ and $y$ coordinates are measured in units of the cluster core
radius $r_c$. Each panel has been obtained by a Montecarlo
simulation using two different seeds (one for the first row and
another for the last row.)}} \label{Stellar distribution}
\end{figure}

Then, if $d_\star$ is the average distance between two close stars
and $\lambda_{GW}$ is the GW wavelength, the Bragg formula gives
the position of the first diffraction peak
\begin{equation}\label{Bragg}
2d_\star\sin\theta=\lambda_{GW}.
\end{equation}
Assuming that the GW source is a binary system on circular orbit,
we obtain the following condition which the cluster physical
parameters should satisfy, in order to have substantial GW
diffraction
\begin{equation}\label{Lambda condition}
0\leq
1-\left(\frac{P_0}{P}\right)^3\left[1+\left(\frac{r}{r_c}\right)^2
\right]^{\frac{5}{2}}\lesssim
f_{BH}<1
\end{equation}
where, $r$ is the distance from the cluster center, $r_c$ is the
cluster core radius, $f_{BH}$ is the black hole mass fraction to
the total cluster mass $M$ and $P_0$ is a ``critical'' period
defined by
\begin{equation}\label{Critical period}
P_0=\frac{4}{c}\left(\frac{4}{3}\pi\right)^{\frac{1}{3}}
\left(\frac{M}{M_\odot}\right)^{-\frac{1}{3}}r_c.
\end{equation}
The physical meaning of the parameter introduced above is the
following: it simply gives the upper and lower limit for the
period of the GW source, in order to get diffraction in the
cluster central region ($r\ut< r_c$). So, in terms of $P$, the
equation (\ref{Critical period}) becomes
\begin{equation}\label{Period condition}
P_0\left[1+\left(\frac{r}{r_c}\right)^2\right]^{\frac{5}{6}}\leq
P\lesssim
P_0(1-f_{BH})^{-\frac{1}{3}}\left[1+\left(\frac{r}{r_c}\right)^2\right]^{\frac{5}{6}}.
\end{equation}
For example, considering the cluster at the center of our Galaxy
(for which we have $M=3.67\cdot 10^6~M_\odot$, $r_c=5.8\cdot
10^{-3}~{\rm pc}$, $R=100~r_c$), we finally get $P_0=2.4\cdot
10^4~{\rm s}=6.7~{\rm h}$.


\subsection{Diffraction patterns}

Treating the problem in the Fraunhofer approximation, the
intensity distribution on a plane screen is given by \citep{born}
\begin{equation}\label{Intensity}
I(x,y)=I_0(x,y)\sum_{m=1}^{N}\sum_{n=1}^{N}e^{ik\left((\xi_m-\xi_n)\frac{x}{F}+(\eta_m-\eta_n)\frac{y}{F}\right)}
\end{equation}
where $I_0(x,y)$ is the diffraction pattern of a reference slit
and the double sum takes into account the interference between
each slit, $(\xi_m,\eta_m)$ are the coordinates of the $m$-th
slit, $N$ is the total number of slits, $F$ is the distance of the
cluster plane from the screen plane, and $k$ is the GW wavenumber.
For a randomly distributed slit sample, relation (\ref{Intensity})
becomes
\begin{equation}\label{Intensity (randomly)}
I(x,y)\simeq NI_0(x,y).
\end{equation}
Since we cannot consider a random distribution of stars, we must
use the more general equation with a star distribution weighted by
the Plummer function.

\begin{figure}[h]
\begin{center}
\hbox{\hspace{3.5cm}
\psfig{figure=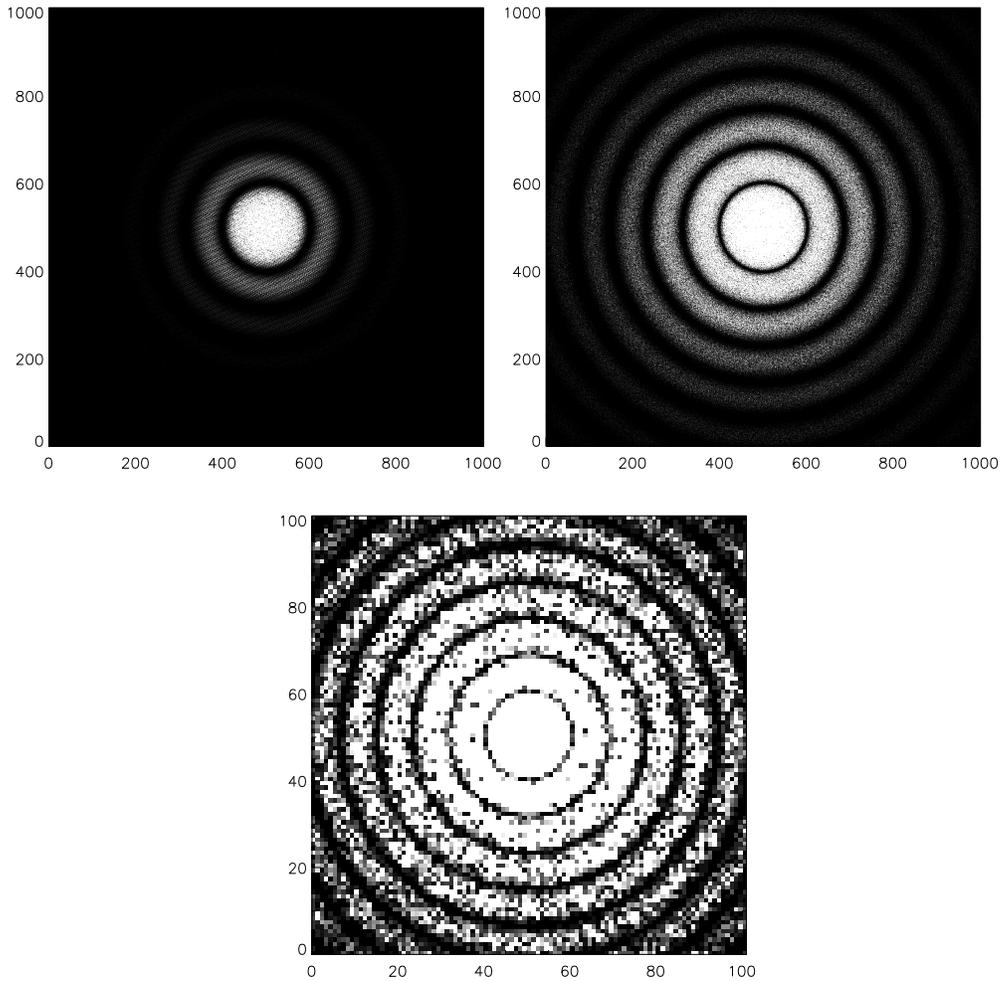,height=7cm,width=9.7cm}
} \vspace{-1.0cm}
\end{center}
\caption{\footnotesize {Each panel shows the expected diffraction
pattern for $N=10,~100$ and $1000$ stars, respectively. Here, the
spatial resolution is $(1000~{\rm pxl})^2 = (5.1~{\rm kpc})^2$ for
the two upper images, while it is $(100~{\rm pxl})^2=(5.1~{\rm
kpc})^2$ for the third image (on the bottom).}} \label{Pattern}
\end{figure}

Now, by considering a star cluster at the galactic center with
mass $M=3.67\cdot 10^6~M_\odot$, $N=10,~100,~1000$ stars,
$\lambda_{GW}=1.5\cdot 10^{16}~{\rm cm}$, $d_\star=1.5\cdot
10^{17}~{\rm cm}$, $F=8.5~{\rm kpc}$ (Earth distance from the
galactic center), we obtain the following diffraction patterns
(see Figure \ref{Pattern}).

\begin{figure}[h]
\begin{center}
\hbox{\hspace{0cm}
\psfig{figure=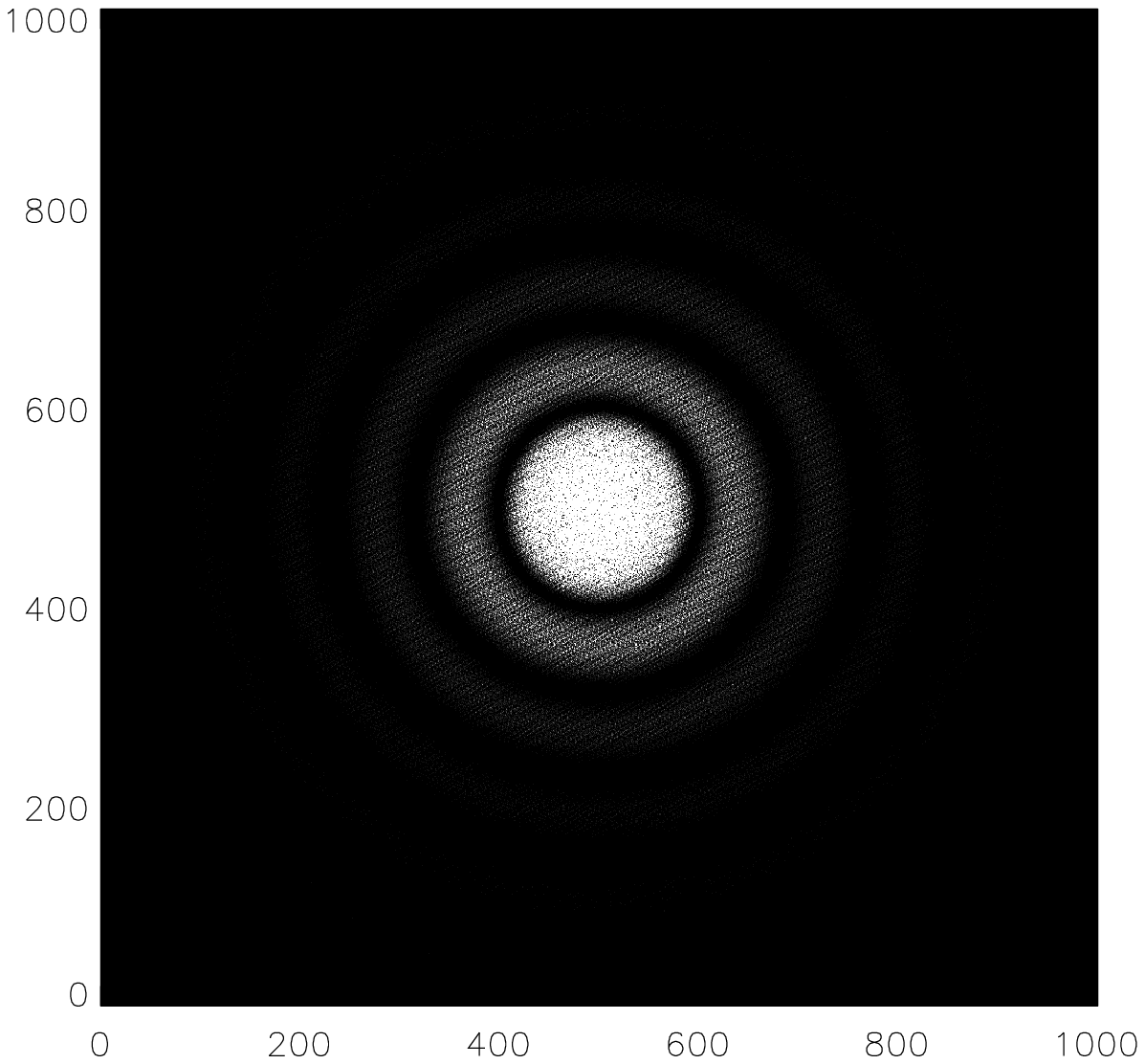,height=7cm,width=9.7cm}
\hspace{-3.2cm}
\psfig{figure=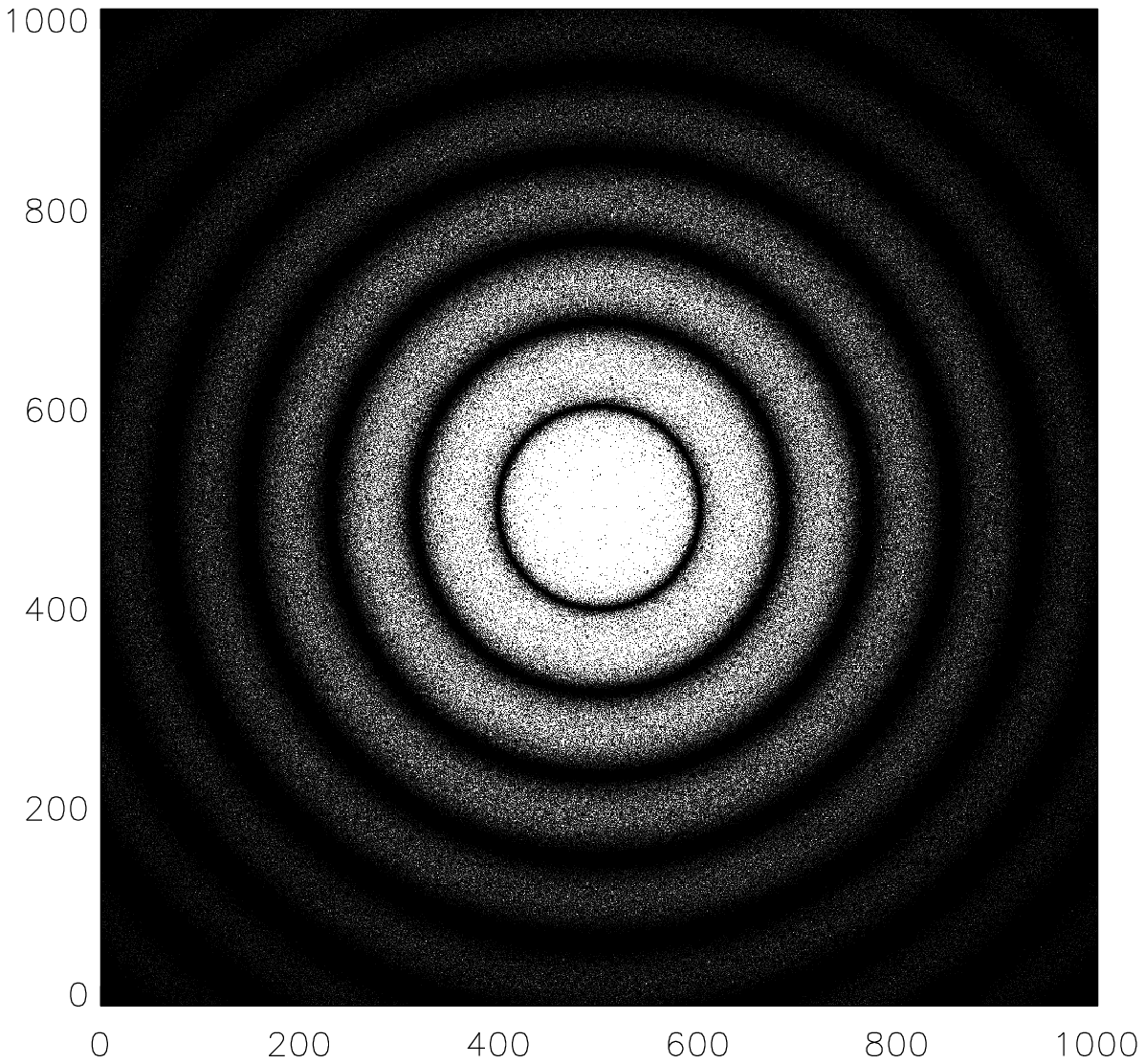,height=7cm,width=9.7cm}
} \vspace{-1.8cm}
\end{center}
\end{figure}

As evident from Figure \ref{Pattern}, each pattern is very similar
to that of a single circular slit of radius $r_c$. Hence, the
Rayleigh criterion allows us to estimate the width of the first
diffraction peak as
\begin{equation}\label{Rayleigh criterion}
\delta\theta=1.22\frac{\lambda_{GW}}{r_c}.
\end{equation}
Using the above values for the cluster parameters, we get
$\delta\theta=58$ degrees, that is $F\delta\theta=8.7~{\rm kpc}$.
On the basis of this calculation, Earth will take about $42\cdot
10^6$ years to cross the whole diffraction maximum, due to its
movement around the galactic center. To obtain this result we have
used an orbital velocity of about $2\cdot 10^7~{\rm cm}$ $s^{-1}$
and assumed that the orbit is approximately straight in the area
included in the angle $\delta\theta$. Because of the ellipticity
of the orbit, it is clear that Earth's crossing time must be
slightly shorter than $42\cdot 10^6$ years.



\begin{thebibliography}{}

\bibitem[\protect\citeauthoryear{Born J. and Wolf S.}{1999}]{born}
Born J. and Wolf S., 1999, Principles of Optics, Cambridge
University Press

\bibitem[\protect\citeauthoryear{Bini et al.}{2005}]{bini2005}
Bini D., De Paolis F., Geralico A. et al., 2005, Gen. Rel. Grav.,
37, 1263

\bibitem[\protect\citeauthoryear{De Paolis et al.}{2002}]{depaolis}
De Paolis et al., 2002, A\&A 366, 1065

\bibitem[\protect\citeauthoryear{De Paolis et al.}{2003}]{depaoliss2}
De Paolis et al., 2003, A\&A, 409, 809

\bibitem[\protect\citeauthoryear{De Paolis et al.}{2005}]{depaolis2005}
De Paolis et al., 2005, submitted to A\&A

\bibitem[\protect\citeauthoryear{Genzel et al.}{2003}]{Genzel03}
Genzel, R., R. Sch\"odel, Ott T, et al.,  2003, Nature, 425, 934.

\bibitem[\protect\citeauthoryear{Ghez et al.}{2003}]{Ghez03}
Ghez, A.M., Duch$\hat{\rm{e}}$ne G., Matthews K, et al.,  2003,
ApJL, 586, L127.

\bibitem[\protect\citeauthoryear{Ghez et al.}{2004}]{Ghez04}
Ghez, A.M., et al.,  2004, ApJL, 601, L159.

\bibitem[\protect\citeauthoryear{Ghez et al.}{2005}]{Ghez05}
Ghez, A.M., et al.,  2005, ApJ, 620, 744.

\bibitem[\protect\citeauthoryear{Hulse and Taylor}{1975}]{hulse}
Hulse R.A., Taylor J. H., 1975, ApJ, 195, L51

\bibitem[\protect\citeauthoryear{Lommen and Backer}{2001}]{lom}
Lommen A. N. and Backer D. C., 2001, ApJ 562, 297

\bibitem[\protect\citeauthoryear{Peters and Mathews}{1963}]{pet}
Peters P. C. and Mathews J., 1963, PhysRev 131, 435

\bibitem[\protect\citeauthoryear{Ruffa}{1999}]{ruffa}
Ruffa A., 1999, ApJ 517, L31

\bibitem[\protect\citeauthoryear{Sch\"odel et
al.}{2003}]{Schoedel03} Sch\"odel, R., et al.,  2003, ApJ, 596,
1015.

\bibitem[\protect\citeauthoryear{Zakharov et al.}{2005}]{miraggi}
Zakharov et al., 2005, New Astronomy, 10, 479


\end{thebibliography}
\end{document}